\documentclass[amssymb,amsmath,pre,amsfonts,superscriptaddress,twocolumn]{revtex4-1}
\usepackage{graphicx}
\usepackage{color} 
\definecolor{bleu}{rgb}{0,0,0}

\usepackage{bm} 
\usepackage{hyperref} 
\usepackage[caption=false]{subfig}
\usepackage{tabularx}
\usepackage{color}

\begin{document}
%%%%%%%%%%%%%%%%%%%%%%%%%%%%%%%%%%%%%%%/\
\title{Diffusion, subdiffusion and localisation of active colloids in random post lattices}
\author{Alexandre Morin}
\email{alexandre.morin@ens-lyon.fr}
\author{David Lopes Cardozo}
\email{david.lopes\_cardozo@ens-lyon.fr}
\author{Vijayakumar Chikkadi}
\author{Denis Bartolo}
\email{denis.bartolo@ens-lyon.fr}
\affiliation{Univ. Lyon, Ens de Lyon, Univ. Claude Bernard, CNRS, 
Laboratoire de Physique, F-69342 Lyon, France}
%%%%%%%%%%%%%%%%%%%%%%%%%%%%%%%%%%%%%%%/\
\begin{abstract}
Combining experiments and theory, we address the dynamics of self-propelled particles in crowded environments.  
We first demonstrate that motile colloids cruising at constant speed through random lattices undergo a smooth transition from diffusive, to  subdiffusive, to localized dynamics upon increasing the obstacle density. We then elucidate the nature of these transitions by performing extensive simulations constructed from a detailed analysis of the colloid-obstacle interactions. 
 We evidence that repulsion at a distance and hard-core interactions both contribute to slowing down the  long-time diffusion of the colloids. In contrast,  the localization transition stems solely from excluded-volume interactions and occurs at the void-percolation threshold.  Within this critical  scenario, equivalent to that of the random Lorentz gas, genuine asymptotic subdiffusion is found only at the critical  density where the motile particles explore a fractal maze.
\end{abstract}
\maketitle
%
%%%%%%%%%%%%%%%%%%%%%%%%%%%%%%%%%%%%%%%/\
\section{introduction}
From intracellular transport to the motion of living creatures in natural habitats, virtually all instances of active transport at small scales occur in crowded environments, see e.g. ~\cite{TheraulazReview2017,Franosch2013}. These observations  together with potential applications of synthetic active matter, {have  resulted in a surge of interest in  self-propulsion through heterogeneous media~\cite{StarkReview,Review2016}. However, apart from  rare exceptions~\cite{PoonCrystal, Peruani2013,Stark2016}, most studies have focused on the two-body interactions between self-propelled particles and isolated obstacles or walls~\cite{LaugaPowers,StarkReview,Review2016,Dileonardo2015,PalacciShelley,Goldstein,Dileonardo2017}.}

In contrast, a different line of research has been devoted to the dynamics of ballistic tracers and random walkers in extended crowded media, see e.g.~\cite{Georges1990,Review1992,Klafter2000,Avraham2002}. From a theoretical perspective, the gold standard is the Random Lorentz Gas model, where passive tracers move ballistically, or diffuse, through a random lattice of hard-core obstacles~\cite{Lorentz1905}. The salient features of this minimal model have been quantitatively explained,  from transient subdiffusion, to the localization transition occurring at the void percolation threshold, see~\cite{Franosch2006,Franosch2010,Franosch2011,Charbonneau2015,Aharony} and references therein. From an experimental perspective,  the Lorentz localization scenario  has been qualitatively confirmed only very recently using Brownian colloids~\cite{Dullens2013}. However, unlike passive  colloids, self-propelled particles couple to their environment  not only via their position, but also via their intrinsic orientation, which chiefly dictates their active dynamics. As a consequence the interactions between motile bodies and fixed obstacles can result in counterintuitive behaviors such as collision and avoidance at constant speed~\cite{Reynolds86,Couzin,Morin2017}. Considering the dynamics of self-propelled particles steadily moving in random lattices of repelling obstacle, Chepizhko et al found a phenomenology which qualitatively differs from that of the Lorentz gas~\cite{Peruani2013}. Numerical simulations indeed suggest that active particles undergo genuine  subdiffusion as a result of transient  trapping  over a range of obstacle densities while localization was not reported. In any realistic setting both reorientation at a distance and excluded volume would affect the particle trajectories. However, until now no experiment has addressed the localization of self-propelled bodies in crowded environments. We rectify this situation

In this article, we combine quantitative experiments and extensive numerical simulations to elucidate the  dynamics of  self-propelled particles  in disordered lattices.   We first  investigate  the trajectories of non-interacting active colloids moving at constant speed through repelling obstacles.  We  quantitatively demonstrate how disorder hinders their  diffusion and ultimately confines their trajectories to compact regions.
The very nature of this localization transition is then identified by disentangling the contributions of finite-range deflection and hard-core repulsion.  We evidence that both excluded volume and deflection at a distance result in finite-time subdiffusion. However 
at long times,  deflection at a distance merely renormalizes the particle diffusivity while hard-core repulsion results in a  localization transition \`a la Lorentz, from  diffusive to  fully localized behavior. 
%Figure 1 
\begin{figure*}
\includegraphics[width=\linewidth]{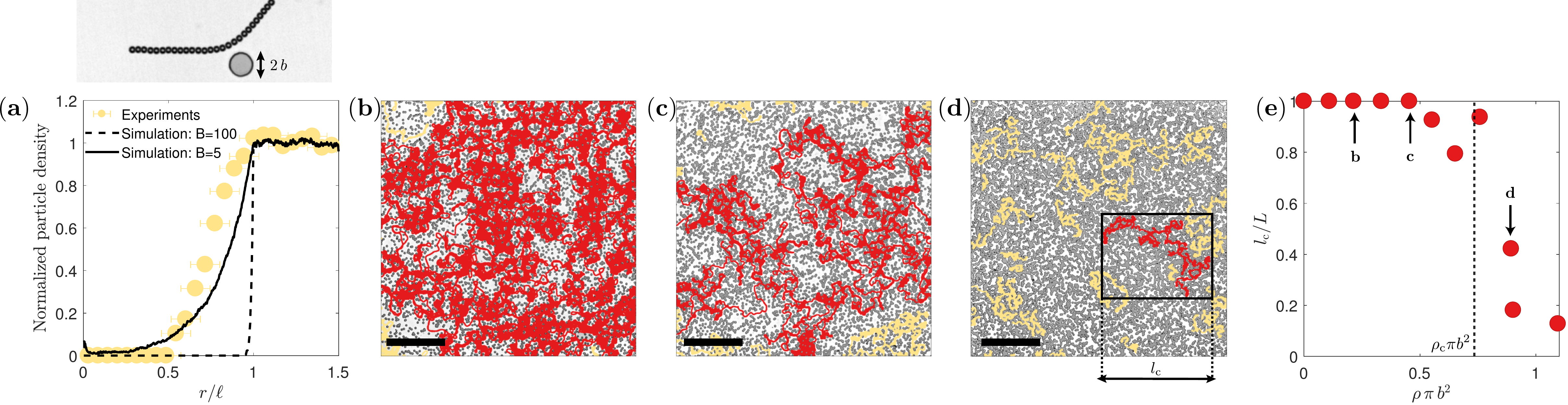}
\caption{Colloidal rollers in random obstacle lattices. (a) Top panel: Superimposed pictures taken at {\em equal time intervals} of a colloidal roller deflected by a lithographied post of radius $b=10\,\rm\mu m$. Note that the direction of motion is changed at constant speed. Bottom panel: 
 Radial density of colloidal rollers propelling around an isolated obstacle. Circles: Experiments. Dark lines: Simulated radial densities. Solid line: $B\ell/v_0=5$, dashed line: $B\ell/v_0=100$ as defined in  \eqref{equationv}.  {\color{bleu}$\ell$ is defined as the value where the density plateaus. In all our experiments we find $\ell\sim 2b$}. {\color{bleu}Error bars: binning size.} (b), (c) and (d)  Trajectories of colloidal rollers (red and yellow) superimposed to the pictures of the obstacle lattices. Scale bar: $500\,\rm\mu m$. Total time: $300\,s$. (b) $\rho= 0.21/(\pi b^2)$, the trajectories form a single percolating cluster. (c)  $\rho = 0.45/(\pi b^2)$, the  trajectories form disconnected clusters. The largest cluster (in red) percolates through the observation region. (d) $\rho = 0.89/(\pi b^2)$, none of the disconnected clusters percolate, and no macroscopic transport is observed. The largest cluster of maximal dimension $l_{\rm c}$ is colored in red. (e) Variations of the normalized maximal cluster size  with the obstacle density. $L$ is the width of the observation window. The dashed line indicates the critical density $\rho_{\rm c}$. {\color{bleu}Experimental errors on the determination of the cluster sizes are smaller than the figure markers. Defining a statistical error on this extremal quantity would require a number of independent realisations beyond our experimental reach.}
}
\label{fig1}
\end{figure*}

\section{Experiments}
\subsection{Exploration of random lattices by colloidal rollers.}
The experimental setup is thoroughly described in Appendix~\ref{Appendix_exp}. Briefly, by taking advantage of the so-called Quincke electro-rotation, we turn polystyrene  beads of radius $a=2.4\,\rm \mu m $ {\color{bleu}immersed in hexadecane (viscosity $\eta \sim 2\,\rm mPa/s$) } into self-propelled colloidal rollers ~\cite{Quincke,Taylor,Smalliuck,Bricard2013}. The basic mechanism of Quincke electro-rotation is recalled in Appendix~\ref{App2}. 
When let to sediment on a flat surface, the colloids roll  at  constant speed $v_{0}=225\,\rm \mu m/s$ along a direction $\hat {\mathbf v}$ which diffuses on the unit circle with an angular diffusivity $D=1.5\,s^{-1}$.  {\color{bleu} Note that thermal diffusion would yield a much lower value of the order of $\sim 5\times10^{-3}\,\rm s^{-1}$. We believe the particle roughness to be chiefly responsible for the spontaneous orientational diffusivity of the rollers.} 
Disorder is introduced by adding UV-lithographied cylindrical posts of  radius $b=10\,\rm \mu m$ on the surface, see Fig.~\ref{fig1} and ~\cite{Morin2017}. The obstacles are placed at random and can overlap. The obstacle density, defined as the number of obstacle centers per unit area, is varied from $\rho=0$ to $\rho=1.1/ (\pi b^2)$.
{\color{bleu} We  focus on a situation opposite to~\cite{Morin2017}, where we considered high roller densities leading to collective flocking motion in  dilute obstacle lattices.} 
 Here, in all experiments, we minimize the interactions between the rollers by keeping their packing fraction far below the onset of collective motion~\cite{Bricard2013,Bricard2015}. In this regime the rollers behave as independent persistent random walkers~\cite{Bricard2013,Bricard2015}.  
We simultaneously track  the trajectories of $\sim100$ colloids in a square observation window of size $L=2.4\,\rm mm$, and all quantities reported below correspond to ensemble and time averages. The  trajectories are recorded at $188$~fps over 5 minutes. During this time interval, particles rolling along straight lines would move over distances of about half a meter.

As illustrated in Fig.~\ref{fig1}a the obstacles repel the rollers at a finite distance while leaving their speed  unchanged. We stress that this behavior is typical of active particles and cannot be observed with passive colloids at thermal equilibrium. The range of the interaction, $\ell\sim 2b$, is measured from the roller density around  isolated obstacles, Fig.~\ref{fig1}a. {\color{bleu}Appendix \ref{App2} provides a detailed analysis of the roller-obstacle interaction (see also the Supplementary Informations of~\cite{Morin2017} for a thorough experimental characterization).} Typical trajectories in random lattices are shown in Figs.~\ref{fig1}b, c, d, and in a Supplementary Video. At low obstacle densities, the rollers freely propagate through the entire system. The ensemble of their trajectories forms a single connected cluster covering most of the free space left around the obstacles. Increasing $\rho$, the trajectories form disconnected and increasingly sparse clusters: a finite fraction of the colloids remains trapped in compact regions.  The extent of the largest cluster is plotted  for all obstacles densities in Fig.~\ref{fig1}e. In agreement with our qualitative observations, above $\rho=0.45/(\pi b^2)$ none of the colloids is observed to cruise through the entire field of view, and the extent of the largest cluster   decreases very sharply at $\rho_{\rm c}\sim 0.75/(\pi b^2)$.\\

%%%FIGURE2
\begin{figure*}
\includegraphics[height=4cm]{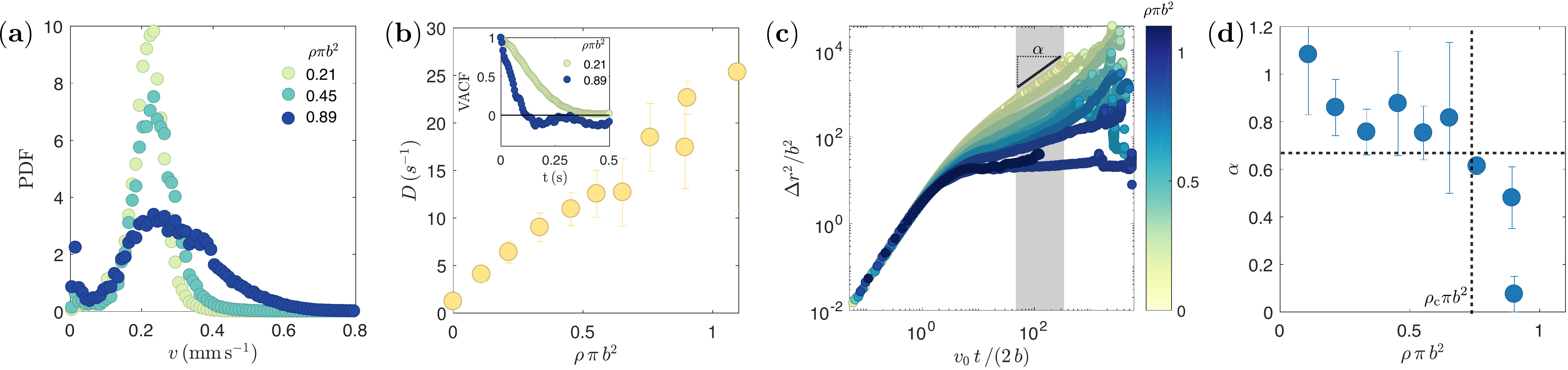}
\caption{
{From active diffusion to localization.} (a) Probability distribution function of the roller speed for three obstacle densities. $\rho= 0.21/(\pi b^2)$, $\rho = 0.45/(\pi b^2)$ and $\rho = 0.89/(\pi b^2)$. The  three distributions are peaked at the same typical speed value: $v_0=250\,\rm \mu m/s$. (b) The angular diffusivity $D$ increases with the obstacle fraction. {\color{bleu} $D$ is measured as the inverse of the time of half decorrelation of the velocity autocorrelation shown in the inset. Error bars: 1 sd.} Inset:  Autocorrelation function of the instantaneous orientation of the rollers velocity ${\rm VACF}=\langle\hat {\mathbf v}(t_0)\cdot\hat {\mathbf v}(t_0+t) \rangle_{t_0}$. (c) Mean squared displacements of the rollers as a function of time. The colors indicate  the obstacle density. The rollers are localized in finite regions at high obstacle densities. (d) Variations of the dynamical exponent $\alpha$ defined as $\Delta r^2\sim t^\alpha$. $\alpha$ is estimated using power-law fits of the mean squared displacements. {\color{bleu} For each obstacle density, independent fits have been performed in 7 intervals of width $v_0 t/(2b)=50$ in the shaded region in (c). $\alpha$ represents the mean of the fitted exponents and error bars represents one $\sigma$.} The vertical dashed line indicates the value of the critical density $\rho_{\rm c}$ defined in the last section. The horizontal dashed lines indicates the value $\alpha=0.66$ corresponding to an ideal overlapping Lorentz gas.}
\label{fig2}
\end{figure*}
%%%

\subsection{\bf Localization of colloidal-roller trajectories.}
The obstacles clearly hinder the exploration of space by the active colloids. However, unlike the situation theoretically considered in~\cite{Stark2016},  the rollers do {\em not}  behave as Active Brownian Particles. They rarely contact the obstacles, and are not slowed down by the collisions. The distribution of their instantaneous speed is peaked at the same value{\color{black}, $v_0=250\,\rm \mu m/s$,} for all obstacle densities, Fig.~\ref{fig2}a. {\color{black}Even more surprisingly, the distribution broadens towards high speeds as $\rho$ increases. This observation alone would imply a faster exploration of space at high obstacle  densities in obvious contrast with our experimental observation, Figs.~\ref{fig1}b, c, and d.}  We  therefore conclude that disorder predominantly impedes the motion of the rollers  by altering their orientational dynamics.  

In Figs.~\ref{fig2}b, we plot the roller orientational diffusivity  $D$, defined as the inverse of the velocity decorrelation time, Fig.~\ref{fig2}b inset. $D$ increases linearly with $\rho$. This  scaling  {\color{black} is} expected for uncorrelated collisions with scatterers all contributing identically to the deflection of the roller trajectories.  Within this  simple  picture  the reduction of the cluster size would merely translate {\color{black} the algebraic decay of the translational diffusivity: $D_{\rm T}\sim v_0^2/D$}, see e.g.~\cite{KirstenLorentz,Review2016,MarchettiFilyReveiw}.  However, the inspection of the mean squared displacements, $\Delta r^2,$ in Fig.~\ref{fig2}c  invalidates this hypothesis. 

At small times, the colloids undergo ballistic motion, however we do not find a universal scaling of the MSDs at long times. The growth exponent $\alpha$ defined as $\Delta r^2\sim t^\alpha$ is a decreasing function of the obstacle density, Fig.~\ref{fig2}d.  
Increasing $\rho$ from 0 to $\rho_{\rm c}$, the long-time dynamics smoothly evolves from normal diffusion ($\alpha=1$) to subdiffusion ($\alpha<1$). Above $\rho_{\rm c}$ the  dynamics slows down abruptly and the rollers undergo a localization transition ($\alpha=0$). The rollers propelling at constant speed, this rich behavior is necessarily encoded in the long-time decay of the  orientational correlations, and therefore cannot be captured by a mere description in terms of an effective  orientational diffusivity~\cite{Franosch2007,Moore1985}.

The central question we aim at answering now is whether the continuous evolution from normal diffusion, to subdiffusion, to  localization, is an asymptotic behavior or a finite-time trend.  Recent simulations of active particles ignoring excluded volume contributions indicate that finite-range repulsion bend the trajectories to form long-live closed orbits. This dynamical trapping results in genuine asymptotic subdiffusion ~\cite{Peruani2013}. In contrast,  within the geometrical picture of the Lorentz gas, subdiffusion should be only observed over finite time scales  diverging only at a critical obstacle fraction $\phi_{\rm L}$~\footnote{Here $\phi_{\rm L}$ is defined as $\phi_{\rm L}=\rho_{\rm L}\pi b^2$, where $b$ in the hard-core radius and $\rho_{\rm L}$, the critical number density for obstacles of radius $b$. Note that $\phi_{\rm L}$ is not the area covering fraction which is given by $1-\exp(-\phi_{\rm L})$}. At $\phi_{\rm L}$, the asymptotic value of $\alpha$ would discontinuously  jump from $1$ to $0$ thereby reflecting a transition toward a fully localized dynamics~\cite{Franosch2006,Franosch2010,Stark2016}.

Clear anticorrelations typical of trapped trajectories are seen in Fig.~\ref{fig2}b inset, yet they are not sufficient to distinguish between the two possible scenarios~\cite{Franosch2007}.  Elucidating the exact nature of the localization transition requires accessing much longer time-scales out of range of our experiments. We resolve this situation by confronting our findings to extensive numerical simulations.

\begin{figure}[b]
\begin{center}
\includegraphics[width=0.5\columnwidth]{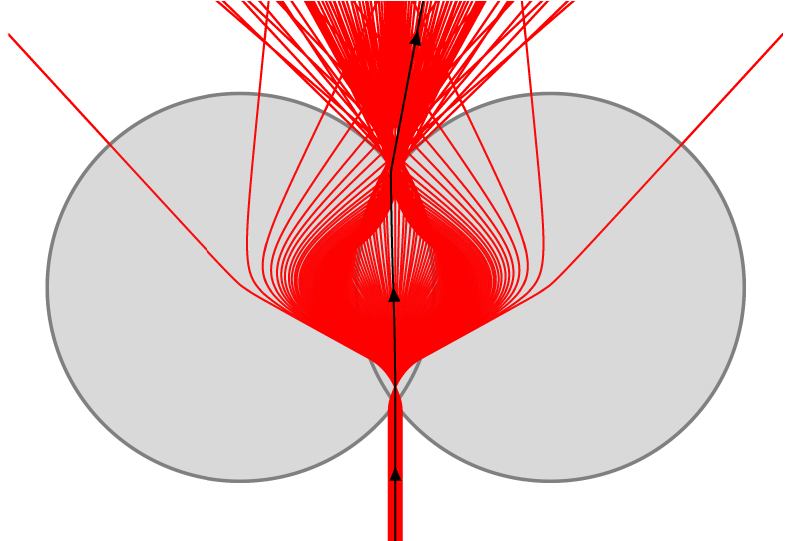}
\caption{Channeling through the obstacles. Ensemble of numerical particle trajectories crossing two obstacles through a narrow channel. $B\ell/v_0=5$. The initial velocity is transverse to the line joining the centers of the two discs. The black trajectory indicates the direction of propagation. Note that particle trajectories nearly equidistant to the obstacles are hardly deflected. All trajectories are spanned at constant speed.
}
\label{Interaction}
\end{center}
\end{figure}

\section{Numerical simulations}
\subsection{Roller-obstacle interactions}

\begin{figure*}[t]
\begin{center}
\includegraphics[height=4cm]{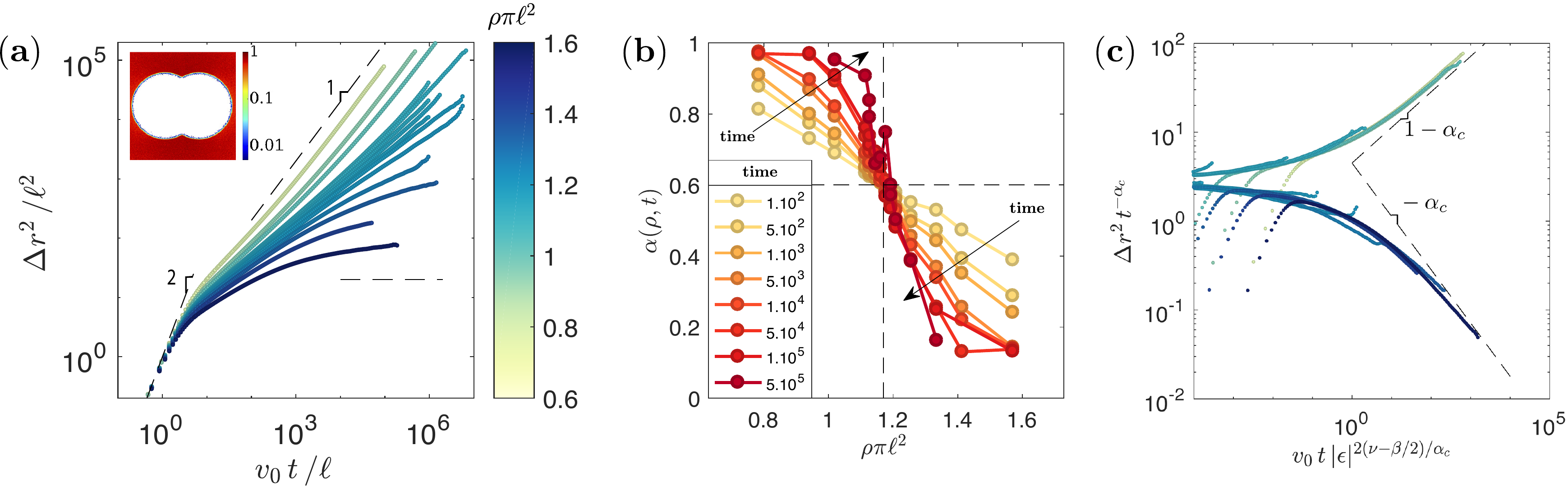}
\caption{Localization transition in the large repulsion limit: $B\ell/v_0=100$. (a) Numerical mean squared displacements of the active particles. The color codes for the obstacle density. Dashed lines correspond to power laws with exponents 2, 1 and 0.
Inset: normalized density map of the active particles around two overlapping interaction discs (log-scale histogram). 
(b) Instantaneous dynamical exponent plotted versus the obstacle density. The different colors correspond to measurements of $\alpha$ at increasing times (expressed in unit of $\ell/v_0$). The dashed lines locate the localization transition.
(c) Scaled MSDs. $\rho_{c}$ and $\alpha_{\rm c}$ are measured from (b) and $\epsilon = (\rho-\rho_c)/\rho_c$. $\beta$ and $\nu$ correspond to the classical percolation exponents. The theoretical values for the overlapping Lorentz gas yield $2(\nu-\beta/2)/\alpha_{\rm c}=4.2$. The best collapse is obtained for $2(\nu-\beta/2)/\alpha_{\rm c}=4.5$. This  discrepancy is very likely to stem from the finite penetration in the obstacles.
}
\label{fig3}
\end{center}
\end{figure*}

Let us first build a simplified phenomenological description of the roller dynamics. {\color{bleu} Details on the numerical resolution of this model are given in Appendix~\ref{Appendix_exp}.}
We need to capture  three central features: (i) the obstacles repel the active colloids isotropically, (ii) the interaction range is finite, (iii) collisions consist in reorientations at constant speed. We also discard spontaneous angular diffusion as it yields minute corrections to the obstacle scattering contributions as $\rho\pi b^2>0.1$, see Fig.~\ref{fig2}b.
Assuming pairwise additive interactions, these observations are sufficient to introduce a general form for the equations of motion of both the roller position $\mathbf r$ and orientation $\hat{\mathbf v}=(\cos\theta,\sin\theta)$:
\begin{align}
\partial_t \mathbf r &= v_0 \mathbf{\hat{v}}(\theta) 
\label{equationr}
\\
\partial_t \theta &= - \partial_\theta \sum\limits_{j} B(\delta r_j) \mathbf{\hat{\delta r_j}} \cdot \mathbf{\hat{v}}
\label{equationv}
\end{align}
where $\mathbf r_j$ is the position of the j$^{\rm th} $ obstacle, and $\mathbf{\delta r_j}  = \mathbf{r_j} - \mathbf{r}$. 
For sake of simplicity $B(\delta r_j)$ is  chosen to be a positive constant, $B$,   for $\delta r_j<\ell$ and 0 otherwise. 
{\color{bleu}We present in Appendix~\ref{App2} a series of experiments complemented by a microscopic theoretical model which ascertains this phenomenological description.}

Before presenting the results of our simulations, let us gain some insight into the roller-obstacles scattering.  
\eqref{equationr} reflects motion at constant speed along $\hat{\mathbf v}$. \eqref{equationv} has a  simple meaning: the rollers turn their back to the obstacles in a typical time $B^{-1}$. In agreement with the trajectory shown in Fig.~\ref{fig1}a, a roller interacting with an obstacle experiences a  torque which orients its velocity in the direction opposite to the vector connecting the roller to the obstacle center.   One important comment is in order. The repelling torques cannot fully exclude  the active particles from the interaction regions.  %Take the simplest possible case of a single roller colliding an isolated obstacle
Take for instance  two obstacles with overlapping interaction disks. 
In the overlap region, the two repulsive torques compete to bend the particle trajectory in opposite directions. As a result, there always exist a finite channel between the obstacles through which the particle can almost freely proceed as illustrated in Fig.~\ref{Interaction}.
%From the same symmetry consideration, ~\eqref{equationv} implies that similar yet narrower channels allow rollers to pass isolated obstacle upon frontal collisions.  
Such interaction-free channels would not exist if the particles were repelled by an isotropic  force (as opposed to an isotropic torque).

%%%%%%%%%%%%%%%%%%%%%%%%%%%%%%%%%%%%%%%

%%%
\subsection{Strong Repulsion Torque and Overlapping Lorentz Gas}
\label{Bgrand}
The particle dynamics is parametrized by a single dimensionless number that compares the time spent in the vicinity of an obstacle ($\ell/v_0$), and the reorientation time $B^{-1}$. In order to see whether repulsion torques can yield subdiffusion and localization as  observed in our experiments, 
it is worth analysing first the  asymptotic case where $B\ell/v_0\gg1$. 
The MSD corresponding to $B\ell/v_0=100$ are plotted in Fig.~\ref{fig3}a. At short times (short distances), the dynamics is ballistic. At intermediate time scales, as in the experiments, we observe a continuous slowing down of the dynamics in the form of subdiffusion as $\rho$ increases. However a careful inspection of the long-time dynamics reveals that this apparent subdiffusion  is merely a transient behavior.  Fig.~\ref{fig3}b shows how the instantaneous value of the exponent $\alpha(\rho,t)$ evolves with time and obstacle density, where $\alpha(\rho,t)=\frac{\rm d}{\rm d\log t}\log\Delta r^2$. The $\alpha(\rho)$ curves converge toward a step function as $t\to\infty$. For $\rho<\rho_{\rm c}=0.3725$, $\alpha(\rho)$ converges to 1. The particles undergo normal diffusion at long times. Conversely, for $\rho>\rho_{\rm c}$, $\alpha\to0$ and  particle motion is localized. As it turns out, the active-particle dynamics is genuinely subdiffusive only at $\rho=\rho_c$ which corresponds to a fixed point of the $\alpha(\rho,t)$ curves. At $\rho_{\rm c}$, $\Delta r^2\sim t^{\alpha_{\rm c}}$ with $\alpha_{\rm c}=0.6\pm0.02$, see Fig.~\ref{fig3}b. Surprisingly,  both the value  of this anomalous exponent and  of $\rho_{\rm c}\pi \ell^2=1.17$  suggest that this localization transition belongs to the  universality class  of the overlapping Lorentz gas model~\cite{Lorentz1905,Franosch2010}. The predictions of the overlapping Lorentz gas would be $\alpha_{\rm c}=0.66$, and $\rho_{\rm c}\pi\ell^2=1.13$. This hypothesis is further confirmed by Fig.~\ref{fig3}c, which shows the collapse of the MSD curves when time and distances are suitably rescaled by the distance to the critical density $\epsilon=(\rho-\rho_{\rm c})/\rho_{\rm c}$ using the Lorentz hyperscaling relations~ \cite{Franosch2010,Franosch2006}.

The universality of the Lorentz localization transition  stems from an underlying  percolation transition~\cite{Aharony}. Localization occurs as the voids separating {\em impenetrable} obstacles stop percolating through the system. In contrast,  
 we  have neglected the hard-core repulsion from the obstacles in ~\eqref{equationr} and ~\eqref{equationv}, and the repelling torques cannot fully exclude  active particles from the interaction regions.  However, in practice, when $B\ell/v_0=100$  the channels allowing the penetration of the  interaction discs become so narrow that we do not observe a single obstacle-crossing event in our simulations, see Fig.~\ref{fig3}a inset. The particles merely penetrate the interaction discs over  minute distances of the order of $v_0/B=\ell/100$ before being strongly repelled. This small yet finite penetration explains the slight discrepancies with the values of the critical density and exponent compared to the ideal Lorentz gaz scenario~\cite{Dullens2013,Horbach2015}. Both $\alpha_{\rm c}$ and $\rho_{\rm c}$ exceed the Lorentz-gas value by $10\%$. The latter difference consistently corresponds to smaller hard-core particles of radius $\ell-2(v_0/B)$.
Is this localization scenario relevant to our experiments? In order to answer this question, we now need testing the robustness of this phenomenology to finite repulsion strength.

\subsection{Finite Repulsion: Diffusion Through Disorder}
%%%%%%%%%%%%%%%%%%%%%%%%%%%%%%%%%
\begin{figure}
\begin{center}
\includegraphics[width=\columnwidth]{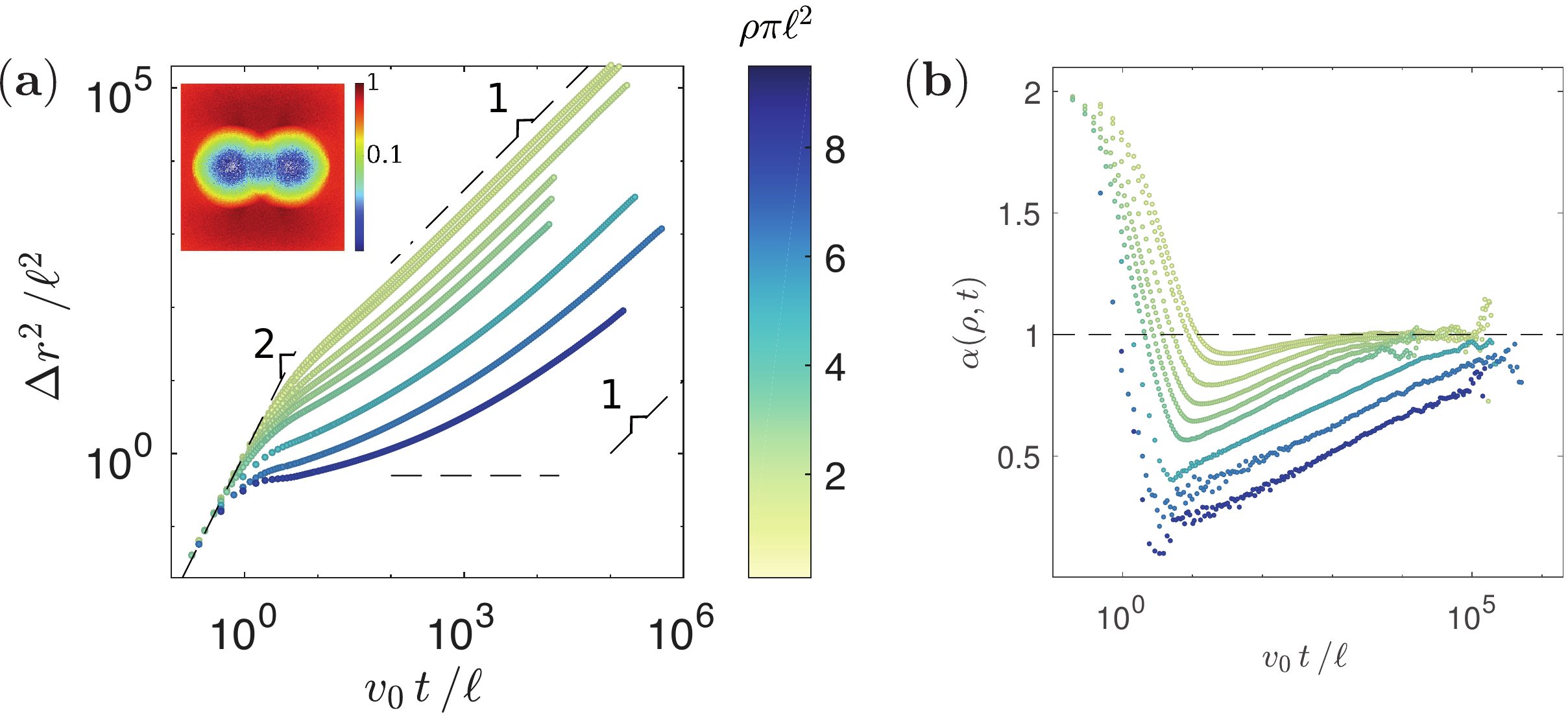}  
\caption{
{ No localization  at finite repulsion: $B\ell/v_0=5$.}
(a) Numerical mean squared displacements of the active particles. The color indicates the obstacle density. The dashed lines correspond to power laws with exponents 2, 1 and 0.
Inset: normalized density  map of the active particles around two overlapping interaction discs (log-scale histogram). 
(b) Time variations of the instantaneous dynamical exponent at different obstacle densities. All exponents converge toward  $\alpha=1$ at long time. The asymptotic dynamics corresponds to normal diffusion. 
}
\label{fig4}
\end{center}
\end{figure}
 The comparison between the numerical and experimental densities around isolated obstacles  indicates that $B\ell/v_0=5$ correctly approximates  the repulsion strength of the lithographied obstacles, see Fig.~\ref{fig1}a.  The MSDs corresponding to $B\ell/v_0=5$ are plotted in Fig.~\ref{fig4}a. Surprinsingly, although  $B\ell/v_0>1$, they show a stark difference with  the strong repulsion limit discussed in the previous section. From $\rho\pi \ell^2=0.95$ to $\rho\pi \ell^2=7.8$, where the  interaction disks cover about  $99.96\%$ of the simulation box, we do not observe any sign of localization. Whereas repulsion still results in subdiffusion at intermediate time scales, the instantaneous dynamical exponent $\alpha$ converges to $1$ at long times even for the highest  obstacle densities, Fig.~\ref{fig4}b. Disorder does not yield asymptotic subdiffusion, and only slows down the rollers motion by reducing their translational diffusivity. 

Repulsion at finite $B$ fails in building effective barriers as illustrated in Fig.~\ref{fig4}a inset. For $B\ell/v_0=5$, we see that the width of the channel going through a pair of obstacles  compares to the inter-obstacle distance thereby preventing any form of long-time trapping. As a result, at long times, neither localization nor  subdiffusion can be achieved as both processes  rely on the formation of traps  with diverging escape times~\cite{Georges1990}. 

\subsection{Origin of the Localization Transition in Colloidal-Roller Experiments}
\label{lastsection}
\begin{figure}
\begin{center}
\includegraphics[width=\columnwidth]{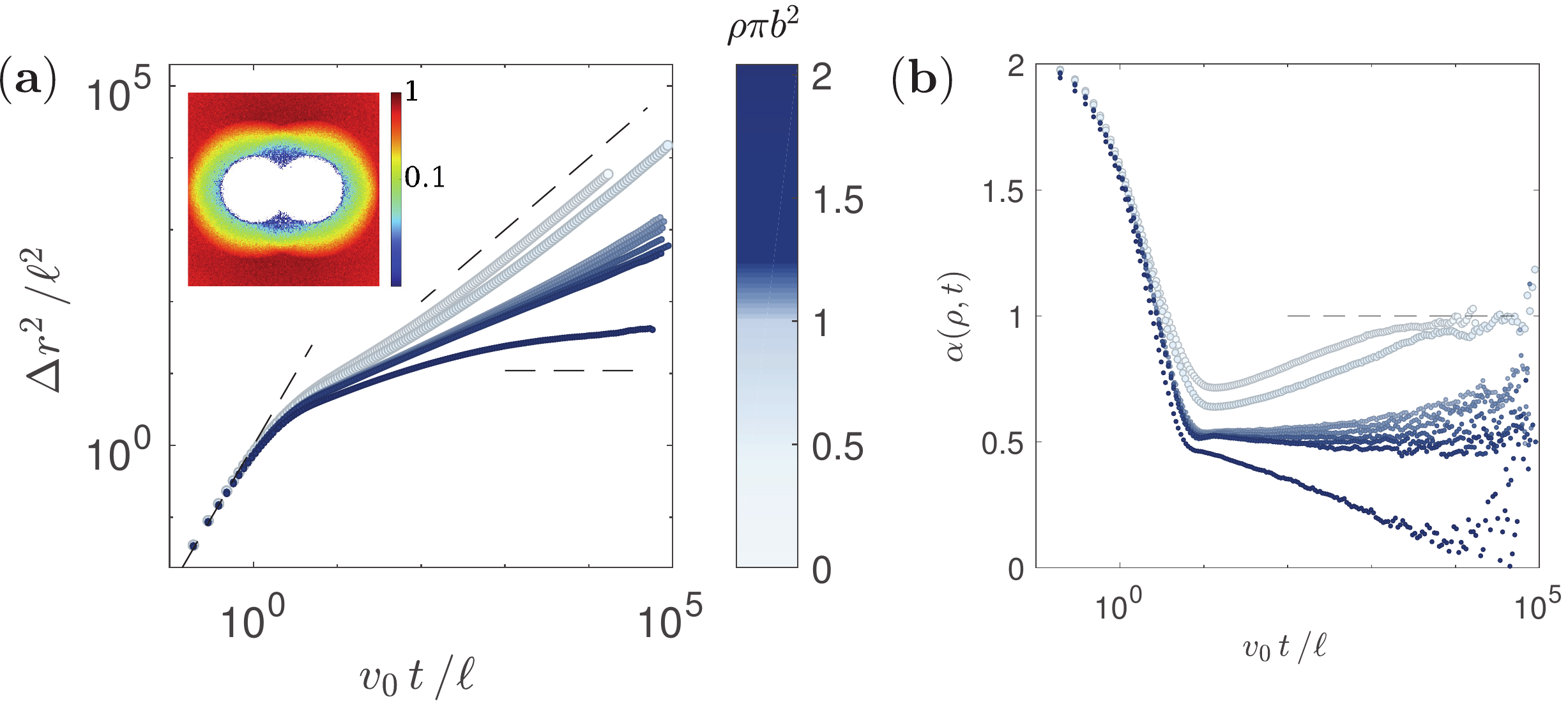}   
\caption{
{Disentangling the roles of hard-core interactions and finite-range repulsion.}
(a)
Numerical mean squared displacements of the active particles. The color indicates the value of $\rho\pi b^2$ and each curve corresponds to a different value of $b$ at constant $\rho=2.05/(\pi \ell^2)$ and $B=5v_0/\ell$. The three dashed lines correspond to power laws with exponents 2, 1 and 0. Inset: Normalized histogram of the roller density (log-scale histogram). 
(b) Time variations of the instantaneous dynamical exponent corresponding to the MSD plotted in (a). Localization occurs only above the percolation threshold of the hard-core obstacles, viz for $\rho\pi b^2>\phi_{\rm L}$. 
}
\label{fig4bis}
\end{center}
\end{figure}
We  infer from the above analysis that the localization transition  must  arise from  the excluded-volume interactions as it  cannot  stem from  hydrodynamic and electrostatic repulsions alone.  In order to test this final hypothesis, we add steric repulsion to the finite-range repulsion torque (keeping $B\ell/v_0=5$). Given the observations reported before, a simple implementation of steric interactions is achieved in adding a repulsion torque of magnitude $B=100v_0/\ell$ and range $b<\ell$ to ~\eqref{equationv}. 

We do recover the experimental phenomenology which turns out to be qualitatively similar to that of the Lorentz model,  Figs.~\ref{fig4bis}a and~\ref{fig4bis}b.  For packing fractions (computed with the hard core radius) smaller than the critical  fraction at the Lorentz transition, $\rho\pi b^2<\phi_L$,  a ballistic regime is followed by a transient subdiffusive dynamics. However, we see that at long times the dynamics ultimately crosses over toward pure diffusion . Approaching $\phi_{L}$ the  extent of the transient regime diverges and yields asymptotic  subdiffusion with $\alpha=0.5\pm0.02$.  
Above $\phi_{\rm L}$ the particles explore  finite regions of space and $\alpha$ converges to 0. The gross features of the dynamics are well captured by a Lorentz scenario, as further confirmed by taking into account the finite size of the rollers when computing the  critical fraction: $\rho_{\rm c}\pi(b+a)^2=\phi_{\rm L}$. This correction gives $\rho_{\rm c}\pi b^2=0.73$ where we expect $0.5<\alpha(\rho_{\rm c})<0.66$ from our simulations. Both values are in excellent agreement with our experimental findings, as shown in Figs.~\ref{fig1}e and ~\ref{fig2}d.

\section{Conclusion}
We have combined quantitative experiments and  extensive simulations to elucidate the dynamics of active particles in random lattices of repelling obstacles. 
We conclude from this  analysis that both repulsion at a distance and excluded volume hinders the exploration of random lattices in the form of transient subdiffusion. 
We show that active colloids cruising through disordered lattices provide a prototypical realization of a random Lorentz gas
undergoing a genuine localization transition at the void percolation threshold. %This conclusion does nots rely on the specifics of the propulsion and and interactions mechanisms and is therefore expected to apply to all microscopic motile bodies exploring crowded environments.

\section*{Acknowledgements}

We acknowledge support from ANR grant MiTra, and Institut Universitaire de France (D.B.). Simulations have been performed at the PSMN computation facility at ENS de Lyon. {\color{black} We thank N. Bain and V. Demery for valuable comments on our manuscript.} 
A. M. and V. C. performed the experiments. D. L. C. performed the numerical simulations. D. B. designed the research. 
A. M., D. L. C. and D. B. discussed the results and wrote the article. 
A. M. and D. L. C. have equally contributed to this research.  

\appendix

\section{Experimental and Numerical Methods}
\label{Appendix_exp}
\subsection{Experiments}
The rollers are fluorescent Polystyrene colloids of diameter $2a=4.8\,\rm \mu m$ dispersed in a 0.11 mol.L$^{-1}$ AOT-hexadecane solution (Thermo scientific G0500). The suspension is injected in a wide microfluidic chamber made of two parallel  glass slides coated by a conducting layer of Indium Tin Oxyde (ITO) (Solems, ITOSOL30, thickness: 80 nm)~\cite{Bricard2013}. The two electrodes are assembled with double-sided scotch tape of homogeneous thickness ($110\,\mu \rm m$). 
The colloids are confined in a $1\,\rm cm\times 1\, cm$ square chambers, by walls  made of a positive photoresist resin (Microposit S1818, thickness: 2 $\mu$m). Identical cylindrical obstacles of radius $b=10\,\mu\rm m$ made of the same material are included in the chambers. Their position is uniformly distributed with a density $\rho$. Therefore the obstacles  can overlap. This geometry is achieved by means of conventional UV lithography. 
 
 The colloids are observed at a 4.8X magnification with a fluorescent Nikon AZ100 microscope. The movies are recorded with a CMOS camera (Basler ACE) at frame rates of 188 fps.  The particles are detected to sub-pixel accuracy, and the particle trajectories and velocities are reconstructed using  the Crocker and Grier  algorithm \cite{Grier} using an improved version of the Blair and Dusfresne MATLAB code.  Measurements are performed in $2.4\, \rm mm\times 2.4\,\rm mm$ observation windows. 
 
The Quincke electro-rotation of the colloids is controlled by applying a homogeneous electric field transverse to the two electrodes $\mathbf E=E_0\hat{\mathbf z}$. The field is applied with a voltage amplifier (TREK 609E-6). All the reported results correspond to an electric field $E_0=1.1E_{\rm Q}$, where $E_{\rm Q}$ is the Quincke electro-rotation threshold $E_{\rm Q}=0.9\, \rm V/\mu m$.

 \subsection{Simulations}
We numerically solve Eqs.~\eqref{equationr} and \eqref{equationv} using a forward Euler integration scheme with an adaptive time step. The time step $\delta t$ is chosen to be $\delta t=10^{-3}/B\times\min[1,1/\sum_{j} \mathbf{\hat{\delta r_j}} \cdot \mathbf{\hat{v}}]$ in Eq.~\ref{equationv}. The summation over the obstacles is performed by first updating the list of the obstacles interacting with each self-propelled particle. $\ell$ and $v_0$ set the  length and time units. Simulations are performed in $200\times200$ or $1000\times1000$ square boxes. The code is parallelised assigning one trajectory to each independent core. Statistics are performed on 320 to 3200 noninteracting particles for a number of independent realizations of  disorder ranging from 1 (low densities) to 128 (high densities). Typical simulations are launched on 32 to 128 independent cores for hours to weeks on Intel E5-2670 sandy bridge octacore $2.60\,\rm GHz$ processors.
%%%%%%
{\color{bleu}
\section{Roller-Obstacle Interactions}
\label{App2}
In this appendix, we first  review the self-propulsion of the colloidal rollers. Then, combining experiments and theory, we explain the response  of the rollers to external electric and  hydrodynamic driving fields. We finally exploit this result to account for the effective repelling interactions with the cylindrical obstacles. 

\subsection{Quincke Motorization}
\label{Quincke_electrorot}
\begin{figure}
\begin{center}
\includegraphics[width=0.8\columnwidth]{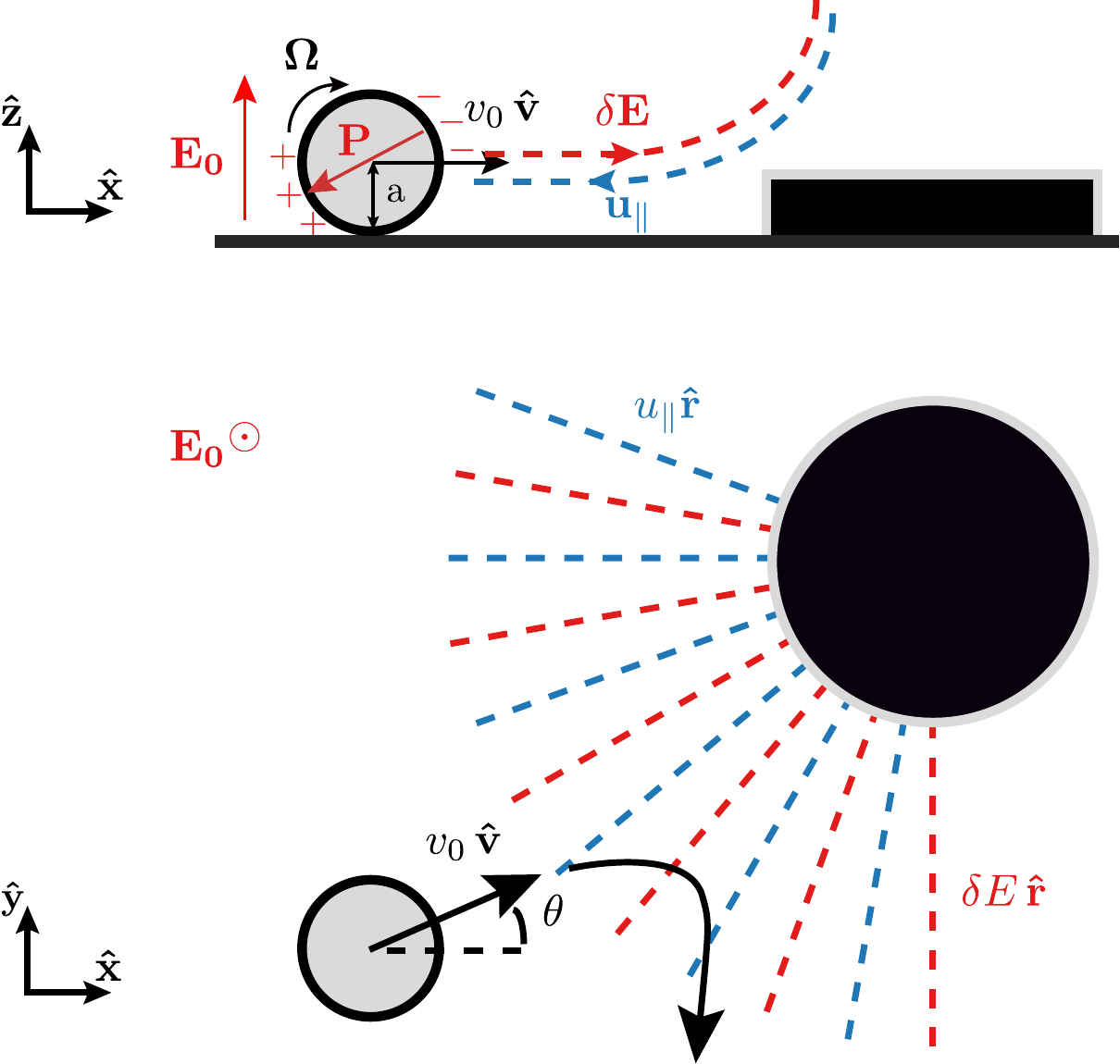}
\end{center}
\caption{Sketch of a colloidal roller propelling near an obstacle. Top panel: Side view. Bottom panel: Top view. The tilted electric dipole at the surface of the roller stems from the Quincke instability. The resulting electric torque drives the roller at constant speed $v_0$. The dielectric obstacle induces  {\em radial} perturbations to the electric field (red dashed lines), and to the fluid-velocity field (blue dashed line).}
\label{fig5}
\end{figure}

The principle of the Quincke motorization is thoroughly discussed in~\cite{Bricard2013} and~\cite{Taylor}. Briefly,  when a homogeneous DC electric field is applied to an insulating sphere  immersed in a conducting fluid, the conduction charges in the solution polarize the solid surface. For fluids and insulating bodies with standard permittivities, the orientation of the resulting electric dipole points in the direction opposite to the electric field.  This situation turns out to be unstable above a critical field amplitude $E_{\rm Q}$. Above $E_{\rm Q}$ any infinitesimal perturbation of the dipole orientation is exponentially amplified. The finite angle made by the electric dipole $\mathbf P$ with the electric field $\mathbf E_0$ results in a net electric torque $\frac{\epsilon_{\rm l}}{\epsilon_0}\mathbf P\times \mathbf{E_{\rm 0}}$, where $\epsilon_{\rm l}$ is the liquid permittivity, Fig.~\ref{fig5}. Ignoring inertia, mechanical equilibrium is reached when the rotational viscous drag acting on the sphere balances the electric torque. Angular momentum conservation then reads $ \eta \bm \Omega=\frac{\epsilon_{\rm l}}{\epsilon_0} \mathbf P\times \mathbf E_{0}$, where $\bm \Omega$ is the angular velocity, and $\eta$ the drag coefficient. Similarly, charge conservation implies the balance between the Ohmic current and 
 the advection of the free charges  by the rotation of the sphere. Together these conservation laws set the rotation speed of the sphere to:
\begin{equation}
\Omega=\Omega_{0}\sqrt{\left(\frac{E_0}{E_Q}\right)^2-1},
\end{equation}
when $E_0>E_{\rm Q}$, and to 0 otherwise.  $\Omega_0$ is the inverse of the so-called Maxwell relaxation time of the free charges~\cite{Taylor}. In our experiments this time scale is typically of the order of 1 ms, which explains the highspeed motion of the colloids. Indeed, when the insulating bead is let to sediment on a solid surface, the above reasoning still applies, and rotation is trivially converted into rolling motion~\cite{Bricard2013}. { Applying an electric field also gives rise to electrophoretic forces that act together with gravity to keep the roller in contact with the bottom electrode. }
As opposed to the colloidal rollers used in~\cite{chaikin} which  undergo stronger slip on the solid surface, in our experiments, the rolling coefficient is close to unity. 

We stress that  Quincke rotation stems from a spontaneous symmetry breaking of the surface-charge distribution. Therefore the direction of rotation is {\em not} prescribed by the external field and can freely rotate around the $\mathbf E_{0}$ axis. 

\subsection{ Dynamical response of Colloidal Rollers to Electric and Flow Fields: Experiments and Theory}
\begin{figure}
\begin{center}
\includegraphics[width=\columnwidth]{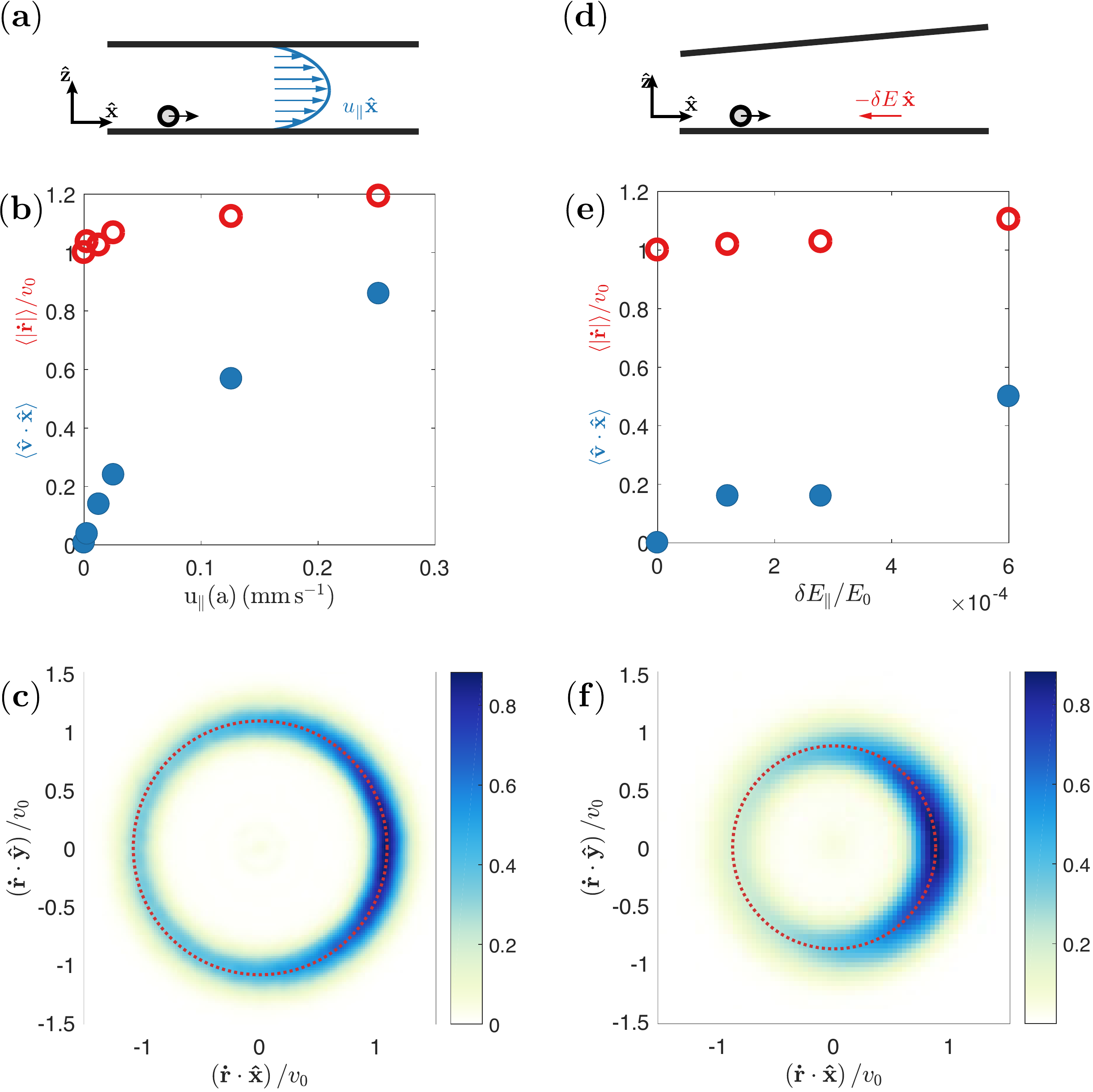}
\end{center}
\caption{(a) Sketch of a colloidal roller driven by a Poiseuille flow. (b) The speed of the roller is hardly modified by the flow (red open symbols), while the projection of the roller velocity on the flow direction $\hat{\mathbf x}$ monotonically increases with the flow (blue filled symbols). (c) PDF of the roller velocity for $u_{\parallel}=0.12\,\rm mm/s$. Note that the distribution remains axisymmetric even at non-zero flow. (d) Sketch of a colloidal roller driven by an electric field. A longitudinal electric field $-\delta E \hat{\mathbf x}$ is applied by tilting the top electrode. (e) As in (b), the speed of the roller is hardly modified (red open symbols) while the projection of the roller velocity on the $\hat{\mathbf x}$-direction  monotonically increases with $\delta E$ evaluated at $z=a$ (blue filled symbols). (f) PDF of the roller velocity for $u_{\parallel}=0.12\,\rm mm/s$. Note again that the distribution remains axisymmetric even at non-zero $\delta E$.  }
\label{fig6}
\end{figure}
In our experiment each obstacle locally alters the direction and the magnitude of the electric field due to their permittivity mismatch with the solvent, and possibly to their net electric charge. In addition, these local heterogeneities of the electric field are very likely to induce electroosmotic flows past the electrodes, see e.g.~\cite{ShraimanNature,NadalEPJE}.  The specifics of the resulting electric and hydrodynamics perturbations goes far beyond the scope of this article; however given the axisymmetric shape of the posts, we can readily infer that both perturbations have a radial symmetry, see Fig.~\ref{fig5}.  The in-plane components of these perturbations can have two consequences on the roller motion: (i) The colloids can be advected by the flow, and, or,  pulled by field gradients. This effect is found to be negligible. In our experiments the roller speed  is hardly modified as they approach the obstacles, Figs.~\ref{fig1}a and \ref{fig2}a. (ii) The rollers can experience external torques which  reorient their electric dipole and velocity, in agreement with the bending of the roller trajectory shown in Fig.~\ref{fig1}a. {Moreover, the out-of- plane component of the electric field increases near the obstacle leading to a broadening of the velocity distribution at high $\rho\pi b^2$, see Fig.~\ref{fig2}a.}

In order to establish a  quantitative description of the roller-obstacle repulsion, we combine theory and dedicated experiments.
In~\cite{Bricard2013}, starting from the Maxwell and Stokes equations we derived the equations of motion of a Quincke roller subject to  a flow field $\mathbf u_{\parallel}$ parallel to the solid surface and to an electric field of the form $\mathbf E=E_0\hat{\mathbf z}+\delta \mathbf E$. They take the simple form:
\begin{align}
&\partial_t{\mathbf r}=v_0\hat{\mathbf v},\label{rpoint}\\
&\partial_t \mathbf{\hat{v}} = (1-\mathbf{\hat{v}}\mathbf{\hat{v}})\cdot \left(\mu_H \partial_z\mathbf{u_\parallel} - \mu_E  {\mathbf{\delta E}}\right),
\label{vpoint}
\end{align}
where $\mu_{H}$ and $\mu_E$ are two positive mobility coefficients, and where both the local shear $\partial_z\mathbf u_{\parallel}$ and the perturbation $\delta \mathbf E$ are evaluated at $z=a$. Given our experimental findings, we  ignore the small corrections to the roller speed that could be caused by transverse perturbations of the electric field and by flow advection (i.e. at $z=a$ we assume $|\mathbf u|\ll v_0$ and $\delta \mathbf E\cdot \mathbf E_0\ll E_0$ ).   

Let us stress that Eqs.~\eqref{rpoint} and \eqref{vpoint} conform to our experimental findings with two  additional experiments. In order to probe the response of colloidal rollers to fluid flows, we apply a Poiseuille flow in an obstacle-free channel, see~Fig.~\ref{fig6}a. We first confirm that the orientational response dominates over  advection: the speed of the rollers, measured in a gas of noninteracting particles,  hardly  increases with the flow (open symbols in Fig.~\ref{fig6}b). In contrast, as the fluid velocity increases we observe that: (i) the orientational distribution is increasingly asymmetric, Fig.~\ref{fig6}c, and (ii) the projection of the average velocity 
 on the flow direction  increases monotonically, filled symbols in Fig.~\ref{fig6}b. The same type of experiment is repeated with electrodes having a wedge   geometry, see  Figs.~\ref{fig6}d. In this geometry, we add a homogeneous longitudinal perturbation to the electric field. Again, the roller speed is unmodified while the angular response is prominent, see Figs.~\ref{fig6}e and~\ref{fig6}f. This set of experiments unambiguously confirm that Eqs.~\eqref{rpoint} and ~\eqref{vpoint} correctly describe the roller dynamics in external driving fields.
\subsection{Effective Interactions with Cylindrical Obstacles}
We finally exploit these results to derive the interaction rules with cylindrical obstacles (Eqs.~\eqref{equationr} and \eqref{equationv} in the main text). Let us consider an obstacle located at the origin. At a point $\mathbf r$ both $\mathbf u_\parallel$ and $\mathbf{\delta E}$ are radial vectors, and therefore Eq.~\ref{vpoint} can then be recast in the form:
\begin{equation}
\partial_t\hat{\mathbf v}=B(r)(1-\mathbf{\hat{v}}\mathbf{\hat{v}})\cdot \hat{\mathbf r},
\label{last}
\end{equation}
where $B(r)=\left(\mu_H \partial_z\mathbf{u_\parallel} - \mu_E  {\mathbf{\delta E}}\right)\cdot\hat{\mathbf r}$. Projecting this equation on the $x$-axis readily yields Eq.~\ref{equationv}. Again the specific expression of $B(r)$ is a complex function of the post shape  and of the material properties.  $B(r)$ is measured to be positive (repulsion) and  to quickly decay with $r$, with a typical range $\ell$ set by the obstacle size, Fig.~\ref{fig1}a. Therefore, for sake of simplicity, we approximate the expression of $B(r)$ by a step function  of width $\ell$. As a final comment we emphasize that Eqs.~\eqref{equationr}~\eqref{equationv}, and~\eqref{last} do not depend on the specifics of the roller-obstacle interactions and hold for any short-range repulsion mechanisms primarily acting on the particle orientation.
}

%%%%%%%%%%%%%%%%%%%%%
%%%%%%%%%%%%%%%%%%%%%

%

\begin{thebibliography}{1}%
\makeatletter
\providecommand \@ifxundefined [1]{%
 \@ifx{#1\undefined}
}%
\providecommand \@ifnum [1]{%
 \ifnum #1\expandafter \@firstoftwo
 \else \expandafter \@secondoftwo
 \fi
}%
\providecommand \@ifx [1]{%
 \ifx #1\expandafter \@firstoftwo
 \else \expandafter \@secondoftwo
 \fi
}%
\providecommand \natexlab [1]{#1}%
\providecommand \enquote  [1]{``#1''}%
\providecommand \bibnamefont  [1]{#1}%
\providecommand \bibfnamefont [1]{#1}%
\providecommand \citenamefont [1]{#1}%
\providecommand \href@noop [0]{\@secondoftwo}%
\providecommand \href [0]{\begingroup \@sanitize@url \@href}%
\providecommand \@href[1]{\@@startlink{#1}\@@href}%
\providecommand \@@href[1]{\endgroup#1\@@endlink}%
\providecommand \@sanitize@url [0]{\catcode `\\12\catcode `\$12\catcode
  `\&12\catcode `\#12\catcode `\^12\catcode `\_12\catcode `\%12\relax}%
\providecommand \@@startlink[1]{}%
\providecommand \@@endlink[0]{}%
\providecommand \url  [0]{\begingroup\@sanitize@url \@url }%
\providecommand \@url [1]{\endgroup\@href {#1}{\urlprefix }}%
\providecommand \urlprefix  [0]{URL }%
\providecommand \Eprint [0]{\href }%
\providecommand \doibase [0]{http://dx.doi.org/}%
\providecommand \selectlanguage [0]{\@gobble}%
\providecommand \bibinfo  [0]{\@secondoftwo}%
\providecommand \bibfield  [0]{\@secondoftwo}%
\providecommand \translation [1]{[#1]}%
\providecommand \BibitemOpen [0]{}%
\providecommand \bibitemStop [0]{}%
\providecommand \bibitemNoStop [0]{.\EOS\space}%
\providecommand \EOS [0]{\spacefactor3000\relax}%
\providecommand \BibitemShut  [1]{\csname bibitem#1\endcsname}%
\let\auto@bib@innerbib\@empty
%</preamble>
\bibitem [{Note1()}]{Note1}%
  \BibitemOpen
  \bibinfo {note} {Here $\phi _{\protect \rm L}$ is defined as $\phi _{\protect
  \rm L}=\rho _{\protect \rm L}\pi b^2$, where $b$ in the hard-core radius and
  $\rho _{\protect \rm L}$, the critical number density for obstacles of radius
  $b$. Note that $\phi _{\protect \rm L}$ is not the area covering fraction
  which is given by $1-\protect \qopname \relax o{exp}(-\phi _{\protect \rm
  L})$}\BibitemShut {NoStop}%
\end{thebibliography}%


\begin{thebibliography}{10}

\bibitem{TheraulazReview2017}
Perna, A., \& Theraulaz, G. When social behaviour is moulded in clay: on growth and form of social insect nests. {\em The Journal of Experimental Biology}, \textbf{220}, 83--91 (2017).

\bibitem{Franosch2013}
H\" ofling, F., \& Franosch, T.  Anomalous transport in the crowded world of biological cells. {\em Rep. Prog. Phys.}, \textbf{76}, 046602 (2013). 



\bibitem{StarkReview}
Z\"ottl, A., \& Stark, H. Emergent behavior in active colloids. {\em J. Phys.: Condens. Matter}, \textbf{28}, 253001 (2016).

\bibitem{Review2016}
Bechinger, C., Di Leonardo, R., L\"owen, H., Reichhardt, C., Volpe, G., \& Volpe, G. Active particles in complex and crowded environments. {\em Rev. Mod. Phys.} \textbf{88}, 045006 (2016). 

\bibitem{PoonCrystal}
Brown, A. T., Vladescu, I. D., Dawson, A., Vissers, T., Schwarz-Linek, J., Lintuvuori, J. S., \& Poon, W. C. K. Swimming in a crystal. {\em Soft Matter}, \textbf{12}, 131--140 (2016).

\bibitem{Peruani2013}
Chepizhko, O. \& Peruani, F. Diffusion, Subdiffusion, and Trapping of active particles in heterogeneous media. {\em Phys. Rev. Lett.} \textbf{111}, 160604 (2013).

\bibitem{Stark2016}
Zeitz, M., Wolff, K., \& Stark, H. Active Brownian particles moving in a random Lorentz gas. {\em Eur. Phys. J. E}, \textbf{40}, 23 (2017).

\bibitem{LaugaPowers}
Lauga, E., \& Powers, T. R., The hydrodynamics of swimming microorganisms. {\em Rep. Prog. Phys.}, \textbf{72}, 96601--36 (2009).

\bibitem{Dileonardo2015}
Sipos, O. and Nagy, K. and Di Leonardo, R. \& Galajda, P. Hydrodynamic trapping of swimming bacteria by convex walls. {\em Phys. Rev. Lett.}, {\textbf 114}, {258104} (2015).

\bibitem{PalacciShelley}
Takagi, D., Palacci, J., Braunschweig, A. B., Shelley, M. J., \& Zhang, J. Hydrodynamic capture of microswimmers into sphere-bound orbits. {\em Soft Matter}, \textbf{I},, 3--8 (2014).

\bibitem{Goldstein}
Drescher, K., Dunkel, J., Cisneros, L. H., Ganguly, S., \& Goldstein, R. E. Fluid dynamics and noise in bacterial cell-cell and cell-surface scattering. {\em Proc. Nat. Acad. Sci. USA}, \textbf{108}, 10940--10945.  (2011). 

\bibitem{Dileonardo2017}
Bianchi, S., Saglimbeni, F., \& Di Leonardo, R. Holographic imaging reveals the mechanism of wall entrapment in swimming bacteria. {\em Phys. Rev. X}, \textbf{7}, 11010 (2017).


\bibitem{Review1992}
Isichenko, M. B. Percolation, statistical topography, and transport in random media. {\em Rev. Mod. Phys.} \textbf{64}, 961 (1992). 

\bibitem{Georges1990}
Bouchaud, J.-P., \& Georges, A. Anomalous diffusion in disordered media: statistical mechanisms, models and physical applications. {\em Physics Reports}, \textbf{195}, 127--293 (1990).

\bibitem{Klafter2000}
Metzler, R., \& Klafter, J. The random walk's guide to anomalous diffusion: a fractional dynamics approach. {\em Physics Reports}, \textbf{339}, 1--77 (200). 

\bibitem{Avraham2002}
Havlin, S. \& Ben-Avraham, D. Diffusion in disordered media, {\em Adv. Phys.}, \textbf{51}, 187--292, (2002).

\bibitem{Lorentz1905} H. A. Lorentz. The motion of electrons in metallic bodies. {\em KNAW, Proceedings}, \textbf{7}, 438--453 (1905).

\bibitem{Franosch2006}
H\" ofling, F., Franosch, T., \& Frey, E. Localization transition of the three-dimensional Lorentz model and continuum percolation. {\em Phys. Rev. Lett.} \textbf{96}, 165901 (2006).

\bibitem{Franosch2010}
Bauer, T. and H\"ofling, F. and Munk, T. and Frey, E., \& Franosch, T. The localization transition of the two-dimensional Lorentz model. {\em Eur. Phys. J. Special Topics} \textbf{189}, 103 (2010).

\bibitem{Franosch2011}
Franosch, T., Spanner, M., Bauer, T., Schr\"oder-Turk, G. E., \& H\"ofling, F. Space-resolved dynamics of a tracer in a disordered solid. {\em Journal of Non-Crystalline Solids}, \textbf{357}, 472--478, (2011).

\bibitem{Aharony}
Stauffer, D., \& Aharony, A. Introduction to percolation theory. {\em Taylor \& Francis}, (1994)

\bibitem{Charbonneau2015}
Jin, Y., \& Charbonneau, P. Dimensional study of the dynamical arrest in a random Lorentz gas. {\em Phys. Rev. E}, \textbf{91}, 042313 (2015).

\bibitem{Dullens2013}
Skinner, T. O. E., Schnyder, S. K., Aarts, D. G. A. L., Horbach, J., \& Dullens, R. P. A. Localization dynamics of fluids in random confinement. {\em Phys. Rev. Lett.} \textbf{111}, 128301 (2013).

\bibitem{Reynolds86}
Reynolds, C. W. Flocks, herds and schools: A distributed behavioral model. {\em Comput. Graph.}, \textbf{21}, 25--34  (1987).

\bibitem{Couzin}
Couzin, I. D., Krause, J., James, R., Ruxton, G. D. \& Franks, N. R. Collective memory and spatial sorting in animal groups. {\em Journal of Theoretical Biology}, \textbf{218}, 1--11 (2002) .

\bibitem{Morin2017} Morin, A., Desreumaux, N., Caussin, J.-B., \& Bartolo, D. Distortion and destruction of colloidal flocks in disordered environments. {\em Nat. Phys.} \textbf{13}, 63--67 (2017).

\bibitem{Bricard2013} Bricard, A., Caussin, J.-B., Desreumaux, N., Dauchot, O. \& Bartolo, D.  Emergence of macroscopic directed motion in populations of motile colloids. {\em Nature} \textbf{503}, 95--98 (2013).

\bibitem{Quincke} Quincke, G. \"Uber Rotationen im constanten electrischen Felde. {\em Ann. Phys.} \textbf{295} 417--486 (1896).

\bibitem{Taylor} Melcher, J. R., \& Taylor, G. I. Electrohydrodynamics: a review of the role of interfacial shear stresses. {\em Ann. Rev. Fluid Mech.}, 111--146 (1969).

\bibitem{Smalliuck} Liao, G., Smalyukh, I. I., Kelly, J. R., Lavrentovich, O. D., \& J\'akli, A. Electrorotation of colloidal particles in liquid crystals. {\em Phys. Rev. E} \textbf{72}, 031704 (2005).

\bibitem{Bricard2015} Bricard, A., Caussin, J.-B., Das, D., Savoie, C., Chikkadi, V.,  Shitara, K.,  Chepizhko, O.,  Peruani, F.,  Saintillan, D., \& Bartolo, D. Emergent vortices in populations of colloidal rollers. {\em Nat. Comm.} \textbf{6}, 7470 (2015).



\bibitem{KirstenLorentz}
Martens, K., Angelani, L., Di Leonardo, R., \& Bocquet, L. Probability distributions for the run-and-tumble bacterial dynamics: an analogy to the Lorentz model. {\em Eur. Phys. J. E}, \textbf{35}, 1--6. (2012).

\bibitem{MarchettiFilyReveiw}
Marchetti, M. C., Fily, Y., Henkes, S., Patch, A., \& Yllanes, D., Minimal model of active colloids highlights the role of mechanical interactions in controlling the emergent behavior of active matter. {\em Curr. Opin. Colloid Interface Sci.}, \textbf{21}, 34--43  (2016).

\bibitem{Franosch2007}
H\" ofling, F., \& Franosch, T. Crossover in the slow decay of dynamic correlations in the Lorentz model. {\em Phys. Rev. Lett.}, \textbf{98}, 140601 (2007). 

\bibitem{Moore1985}
Machta, J., \& Moore, S. M. Diffusion and long-time tails in the overlapping Lorentz gas. {\em Phys. Rev. A}, \textbf{32}, 3164-3167 (1985).

\bibitem{Horbach2015}
Schnyder, S. K., Spanner, M., H\"ofling, F., Franosch, T., \& Horbach, J. Rounding of the localization transition in model porous media. {\em Soft Matter}, \textbf{11}, 701-711 (2015).


\bibitem{Grier}
Crocker, J. C. \&  Grier, G.  Methods of digital video microscopy for colloidal studies. {\em J. Colloid Interface Sci.} \textbf{179} 298--310 (1996).

\bibitem{chaikin}
Driscoll, M., Delmotte, B., Youssef, M., Sacanna, S., Donev, A., \& Chaikin, P. Unstable fronts and motile structures formed by microrollers. {\em Nat. Phys.}, \textbf{1}, 1--6 (2016). 

\bibitem{NadalEPJE}
Nadal, F., Argoul, F., Kestener, P., Pouligny, B., Ybert, C., \& Ajdari, A. Electrically induced flows in the vicinity of a dielectric stripe on a conducting plane. {\em Eur. Phys. J. E}, \textbf{9}), 387--399  (2002).

\bibitem{ShraimanNature}
Yeh, S.-R., Seul, M., \& Shraiman, B. I. Assembly of ordered colloidal aggregrates by electric-field-induced fluid flow. {\em Nature}, \textbf{386}, 57--59 (1997).


\end{thebibliography}
\end{document}